\documentclass{aa}

\usepackage{txfonts}
\usepackage{graphics,epsfig}
\usepackage{graphicx}
\usepackage{amsmath}
\usepackage{amssymb}
\usepackage{rotating}
\usepackage{ulem}
\usepackage{float}
\usepackage{color}
\usepackage{appendix}
\usepackage{caption}

\newcommand{\acc}{\ensuremath{\rm atoms~cm^{-2}}}

\newcommand{\um}{\ensuremath{\rm \mu m}}
\newcommand{\sgas}{$\Sigma_{\rm gas}$}
\newcommand{\shi}{$\Sigma_{\rm gas,atomic}$}
\newcommand{\ssfr}{$\Sigma_{\rm SFR}$}
\newcommand{\sst}{$\Sigma_{*}$}

\title{Extended Schmidt law holds for faint dwarf irregular galaxies}

\titlerunning{Extended Schmidt law in dIrrs}

\author{Sambit Roychowdhury\inst{1}, Jayaram N. Chengalur\inst{2} \and Yong Shi \inst{3,4}}

\institute{Jodrell Bank Centre for Astrophysics, Alan Turing Building, School of Physics \& Astronomy, The University of Manchester, Oxford Road, Manchester M13 9PL, UK.\\
\email{sambit.roychowdhury@manchester.ac.uk}
\and
NCRA-TIFR, Post Bag 3, Ganeshkhind, Pune 411 007, India.
\and
School of Astronomy and Space Science, Nanjing University, Nanjing 210093, China.
\and
Key Laboratory of Modern Astronomy and Astrophysics (Nanjing University), Ministry of Education, Nanjing 210093, China.
}

\authorrunning{S. Roychowdhury, J. N. Chengalur and Y. Shi}

\date{Received; accepted}

\begin{document}



\abstract
{The extended Schmidt law (ESL) is a variant of the Schmidt which relates the surface densities of gas and star formation, with the surface density of stellar mass added as an extra parameter.
Although ESL has been shown to be valid for a wide range of galaxy properties, its validity in low-metallicity galaxies has not been comprehensively tested.
This is important because metallicity affects the crucial atomic-to-molecular transition step in the process of conversion of gas to stars.}
{To empirically investigate for the first time whether low metallicity faint dwarf irregular galaxies (dIrrs) from the local universe follow the extended Schmidt law. Here we consider the `global' law where surface densities are averaged over the galactic discs.
Dwarf irregular galaxies are unique not only because they are at the lowest end of mass and star formation scales for galaxies, but also because they are metal-poor compared to the general population of galaxies.}
{Our sample is drawn from the Faint Irregular Galaxy GMRT Survey (FIGGS) which is the largest survey of atomic hydrogen in such galaxies. 
The gas surface densities are determined using their atomic hydrogen content.
The star formation rates are calculated using {\it GALEX} far ultraviolet fluxes after correcting for dust extinction, whereas the stellar surface densities are calculated using {\it Spitzer} 3.6 $\mu$m fluxes. The surface densities are calculated over the stellar discs defined by the 3.6 $\mu$m images.}
{We find dIrrs indeed follow the extended Schmidt law.
The mean deviation of the FIGGS galaxies from the relation is 0.01 dex, with a scatter around the relation of less than half that seen in the original relation.
In comparison, we also show that the FIGGS galaxies are much more deviant when compared to the `canonical' Kennicutt-Schmidt relation.}
{Our results help strengthen the universality of the extended Schmidt law, especially for galaxies with low metallicities.
We suggest that models of star formation in which feedback from previous generations of stars set the pressure in the interstellar medium and affect ongoing star formation, are promising candidates for explaining the ESL. We also confirm that ESL is an independent relation and not a form of a relation between star formation efficiency and metallicity.}


\maketitle

\section{Introduction}

How gas is converted into stars in galaxies is a fundamental question in galaxy formation and evolution.
One of the ways this question has been sought to be answered is by empirically relating some observable measure of the two quantities; e.g. \citet{s59} proposed that the surface density of the star formation rate (\ssfr) is related to the surface density of gas (\sgas) via a power law: \ssfr~$\propto$~\sgas$^{\rm N}$.
This relation was firmly established on galactic disc-averaged scales for different classes of galaxies ranging from star-forming spirals to circum-nuclear starbursts by \citet{ken98}, into what is known as the canonical Kennicutt-Schmidt (KS) law with N=1.4.
But observational studies over the last two decades have shown that instead of  a unique `law' the coefficient and/or the power law slope of a Schmidt-type relation varies based on whether one is looking at: starburst galaxies at high or low redshifts \citep{dad10}, sub-kpc scales in spiral galaxy discs \citep{big08}, only the molecular gas in spatially resolved regions of spirals \citep{ler13,she14}, very high density gas \citep{gao04}, low surface brightness galaxies \citep{wyd09}, the disc-averaged scales and resolved sub-kpc scales relations in dwarf irregular galaxies \citep{roy09,roy11,roy14, roy15,bol11,shi14}, resolved sub-kpc scales in the outskirts of spirals \citep{big10,roy15}, etc.

One way forward towards defining an universal relation between total gas and star formation rate is to include other parameters in the relation which can affect star formation, based on empirical considerations.
These parameters can help take into account physical processes beyond a simple free-fall in a gas disc of constant height -- which provides the most straightforward explanation of the canonical KS law.
For example, \citet{ryd94} found that \ssfr\ is radially correlated with mass surface density of old stars in nearby spirals.
In \citet{ken98} itself, an alternative relation was investigated which relates the \ssfr\ of normal spirals and starbursts to their disc orbital timescales. 
\citet{boi03} tested modified versions of the `Schmidt law' which included dynamical factors as well as the stellar surface density, on radially averaged measurements from nearby spirals.
\citet{bli04} propounded the hypothesis that the mid-plane hydrostatic pressure of a galactic disc as measured using its stellar surface density, determines the ratio of atomic-to-molecular gas and consequently the SFR.
And in \citet{bli06} they empirically proved the validity of the hypothesis using data from a sample of galaxies spanning a large range in magnitude and metallicity.
\citet{ler08} did a comprehensive study of the radial variation of star formation efficiency (SFE, \ssfr\ per unit \sgas) in a sample of spiral and dwarf galaxies.
They found that simply using free-fall time or orbital time was not enough to explain the variation seen in SFE with radius, and also could not clearly distinguish which of the factors like pressure, stellar density or orbital time scale determined the molecular gas fraction. 
They suggest that physical processes like phase balance in atomic gas, formation and destruction of molecular hydrogen, and stellar feedback, at scales below the sub-kpc resolutions of their observations governs the formation of molecular clouds and hence determines the SFR.

\citet{shi11} (hereafter S11) proposed an extended Schmidt law (ESL) which sought to incorporate the effect of existing stars into the current star formation.
For a wide variety of galaxies spanning low to high redshifts, and ranging from starbursts, spirals, to even the low surface brightness galaxies that deviates significantly from the KS law, S11 found an empirical relation between the surface densities of star formation rate, gas and stellar mass (\sst) in the form of,
\begin{equation}
\frac{SFE}{[yr^{-1}]}~=~\frac{\Sigma_{SFR}}{\Sigma_{gas}}~=~-10^{10.28 \pm 0.08}~\frac{\Sigma_{*}^{\rm 0.48 \pm 0.04}}{[M_{\odot} pc^{-2}]}.
\label{eq:esl}
\end{equation}
S11 also show that the ESL is not a mere recast of the KS law combined with another correlation between the surface densities,
and that an ESL type relation exists at sub-kpc scales for spiral galaxies.
The ESL is therefore a promising candidate for a universal star formation relation.

Star-forming dwarf galaxies occupy a distinct region in the parameter space of galaxy properties especially given their low metallicities, and can therefore be used to test the universal validity of any gas--SFR relation.
Understanding the process of star formation at low metallicities is crucial, as 
the efficiency of atomic-to-molecular gas conversion is very sensitive to the presence of metals in the interstellar medium \citep{kru09a,ste14}.
But this type of galaxies have always been difficult to include in frameworks trying to explain empirical relations between gas and star formation.
For example,
although the observed variations in the KS law with gas tracers for spirals and starbursts could be understood \citep[e.g. see][]{lad15}, the discrepancies with the canonical KS law observed for low mass and low metallicity galaxies could not be easily understood.
Analytical models trying to explain the conversion of gas to stars \citep{kru09,ost10} had to be modified to account for the observations at the low metallicity end \citep{bol11,kru13}, or even new formalisms were proposed to explain the star formation in these galaxies \citep{elm15}.
S11 included data from some low metallicity galaxies in the form of Low Surface Brightness (LSB) galaxies in their ESL relation.
However one crucial missing type which has not been tested against the ESL are the more numerous \citep{kar13} low metallicity star-forming dwarf irregular galaxies (dIrrs), which occupy a parameter space distinct to LSBs as is discussed later in Section~\ref{sec:rnd}.
In this study we empirically check whether faint dIrrs follow the ESL, and discuss the implications of the same.

\section{Sample and method}
\label{sec:samp}

The sample for this study is chosen from the Faint Irregular Galaxy GMRT Survey \citep[FIGGS,][]{beg08}, the largest interferometric survey of H{\sc I} 21 cm emission from dIrrs \footnote{We thank the staff of the GMRT who have made the observations used in this paper possible.
GMRT is run by the National Centre for Radio Astrophysics of the Tata Institute of Fundamental Research.}.
In this study we will use the HI surface density as a measure of the total gas surface density.
This is a reasonable approximation since for the few nearby dwarf galaxies of comparable metallicity for which measurements of CO exist \citep[e.g.][]{bol11,elm13,shi16}, the inferred molecular gas mass is small compared to the mass of the atomic gas. 

In this study we use 3.6 $\mu$m fluxes to determine the stellar mass and FUV emission to measure the star formation rate (SFR).
The commonly used SFR estimators all employ some basic assumptions like a constant SFR over the last $10^8$ years, a standard IMF, etc.
Almost all of the star formation in galaxies occur through localised starbursts, and a constant SFR emerges due to sampling of a large number of such starbursts.
Also at the low SFR characteristic of our sample, stochastic effects become important at the high mass end of the IMF.
Given the above considerations, it can be shown that of the commonly used tracers of SFR H$\alpha$ is much more unreliable as compared to FUV for low SFRs \citep{daS14}.
We therefore arrive at the sample for this work by choosing FIGGS galaxies which have both (i) archival FUV data from {\it GALEX} \footnote{Some of the data presented in this paper were obtained from the Multimission Archive at the Space Telescope Science Institute (MAST). STScI is operated by the Association of Universities for Research in Astronomy, Inc., under NASA contract NAS5-26555. Support for MAST for non-HST data is provided by the NASA Office of Space Science via grant NAG5-7584 and by other grants and contracts.}, and (ii) measured {\it Spitzer} 3.6 $\mu$m fluxes from \citet{dal09}.
The properties of the resultant galaxy sub-sample are listed in Table~\ref{tab:samp}.
Whenever available the gas phase metallicities given in column (7) of the table are based on oxygen abundance measurements in the literature, from either \citet{ber12} or \citet{mar10}.
The latter a compilation of previously reported metallicity measurements.
Otherwise the metallicities listed are estimated indirectly using the luminosity (M$_{\rm B}$) -- metallicity relation for dIs from \citet{ekt10} for the rest of the galaxies.
From Table 1 we can see that our sample dIrrs are local (within 6 Mpc), faint ($M_B$ ranging from $-15$ to $-11$), and of low metallicity ($<$20\% solar). 

\begin{table*}
\caption{The dwarf galaxy sample, with measured values}
\label{tab:samp}
\begin{tabular}{|lccccccccc|}
\hline
Galaxy&M$_{\rm B}$&D&D${\rm{_{HI}}}$&D${\rm{_{Ho}}}$&D${\rm{_{3.6 \mu m}}}$&Z/Z$_{\odot}$&$Log(\Sigma_{HI})$&$Log(\Sigma_{*})$&$Log(\Sigma_{SFR})$\\
      &&(Mpc)&(arcmin)&(arcmin)&(arcmin)&&(M$_{\odot}$ pc$^{-2}$)&(M$_{\odot}$ pc$^{-2}$)&(M$_{\odot}$ yr$^{-1}$ kpc$^{-2}$)\\
&[2]&[3]&[4]&[5]&[6]&[7]&[8]&[9]&[10]\\
\hline
DDO 226          &$-$14.17&4.9~&~3.5&2.24&3.37&0.12~~~~&$-$0.24$^{+0.041}_{-0.045}$&0.11$^{+0.3}_{-0.3}$&$-$4.38$^{+0.18}_{-0.27}$\\
UGC 685          &$-$14.31&4.5~&~3.6&2.40&2.98&0.20$^*$&   0.32$^{+0.041}_{-0.045}$&0.67$^{+0.3}_{-0.3}$&$-$3.83$^{+0.18}_{-0.28}$\\
UGC 4459         &$-$13.37&3.56&~4.5&2.00&2.23&0.13$^*$&   0.97$^{+0.041}_{-0.045}$&0.58$^{+0.3}_{-0.3}$&$-$3.22$^{+0.18}_{-0.29}$\\
UGC 5456         &$-$15.08&5.6~&~2.8&1.62&2.68&0.16~~~~&   0.17$^{+0.041}_{-0.045}$&0.71$^{+0.3}_{-0.3}$&$-$3.26$^{+0.18}_{-0.29}$\\
NGC 3741         &$-$13.13&3.0~&14.6&1.48&3.35&0.09$^*$&   0.71$^{+0.041}_{-0.045}$&0.26$^{+0.3}_{-0.3}$&$-$3.50$^{+0.18}_{-0.29}$\\
CGCG 269$-$049   &$-$13.25&4.9~&~2.6&1.05&1.57&0.06$^*$&$-$0.07$^{+0.041}_{-0.045}$&0.04$^{+0.3}_{-0.3}$&$-$3.88$^{+0.18}_{-0.28}$\\
DDO 125          &$-$14.16&2.5~&~7.0&3.89&4.50&0.19$^{\dagger}$&   0.28$^{+0.041}_{-0.045}$&0.62$^{+0.3}_{-0.3}$&$-$3.80$^{+0.18}_{-0.29}$\\
UGC 7605         &$-$13.53&4.43&~3.3&1.48&2.35&0.10~~~~&   0.34$^{+0.041}_{-0.045}$&0.60$^{+0.3}_{-0.3}$&$-$3.56$^{+0.18}_{-0.28}$\\
GR8              &$-$12.11&2.1~&~4.3&1.66&2.08&0.09$^*$&   0.68$^{+0.041}_{-0.045}$&0.56$^{+0.3}_{-0.3}$&$-$3.21$^{+0.19}_{-0.25}$\\
UGC 8638         &$-$13.68&4.27&~1.2&1.66&2.98&0.18$^{\dagger}$&   0.02$^{+0.041}_{-0.045}$&0.51$^{+0.3}_{-0.3}$&$-$3.73$^{+0.18}_{-0.28}$\\
DDO 181          &$-$13.03&3.1~&~5.2&2.40&3.23&0.14$^*$&   0.32$^{+0.041}_{-0.045}$&0.28$^{+0.3}_{-0.3}$&$-$3.85$^{+0.18}_{-0.28}$\\
DDO 183          &$-$13.17&3.24&~4.6&2.40&3.60&0.09~~~~&   0.10$^{+0.041}_{-0.045}$&0.08$^{+0.3}_{-0.3}$&$-$4.29$^{+0.18}_{-0.28}$\\
UGC 8833         &$-$12.42&3.2~&~3.0&1.17&2.02&0.07~~~~&   0.57$^{+0.041}_{-0.045}$&0.38$^{+0.3}_{-0.3}$&$-$3.63$^{+0.18}_{-0.28}$\\
DDO 187          &$-$12.51&2.5~&~3.4&1.70&2.12&0.11$^*$&   0.77$^{+0.041}_{-0.045}$&0.51$^{+0.3}_{-0.3}$&$-$3.55$^{+0.18}_{-0.28}$\\
KKH 98           &$-$10.78&2.5~&~3.8&1.05&2.10&0.04~~~~&   0.18$^{+0.041}_{-0.045}$&0.31$^{+0.3}_{-0.3}$&$-$4.17$^{+0.19}_{-0.27}$\\
\hline
\end{tabular}
\begin{flushleft}
Notes:
[2] the absolute B-band magnitude \citep[from][]{beg08},
[3] the distance in Mpc \citep[from][]{kar13},
[4] the diameter of the 10$^{\rm -19}$~ \acc\ isophote determined using the coarsest resolution HI map \citep{beg08},
[5] the B-band diameter at 26.5 magnitude arcsecond$^{\rm -2}$ \citep[i.e. the Holmberg diameter from][]{kar13},
[6] the major diameter of the ellipse within which the 3.6 $\mu$m fluxes were determined \citep[from][]{dal09},
[7] the estimated gas phase metallicity (see text for details),
[8] the surface/column density of HI within the `stellar disc' (see text for details),
[9] the surface density of stars within the `stellar disc',
[10] the surface density of star formation rate witin the `stellar disc'.\\
$^*$: Based on abundances from \citet{mar10} using 12+log$_{10}$[O/H]$_{\odot}$~=~8.7.\\
$^{\dagger}$: Based on average abundances from \citet{ber12} using 12+log$_{10}$[O/H]$_{\odot}$~=~8.7.
\end{flushleft}
\end{table*}

\subsection{Estimates of gas and SFR surface densities}

\begin{figure*}
\nopagebreak
\begin{center}
\begin{tabular}{cccc}
DDO 226&UGC 685&UGC 4459&UGC 5456\\
{\mbox{\includegraphics[height=4cm,angle=270]{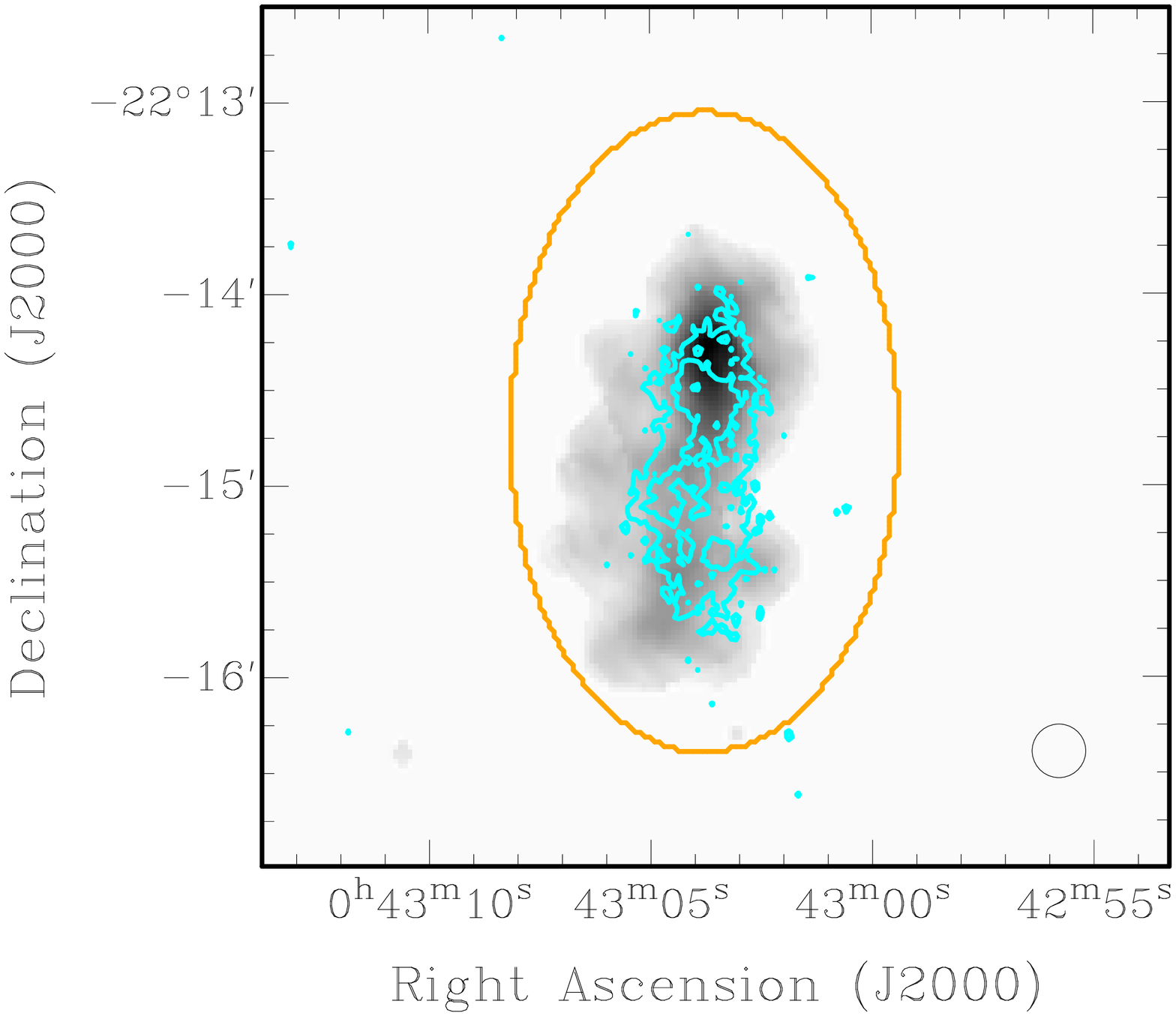}}}&
{\mbox{\includegraphics[height=4cm,angle=270]{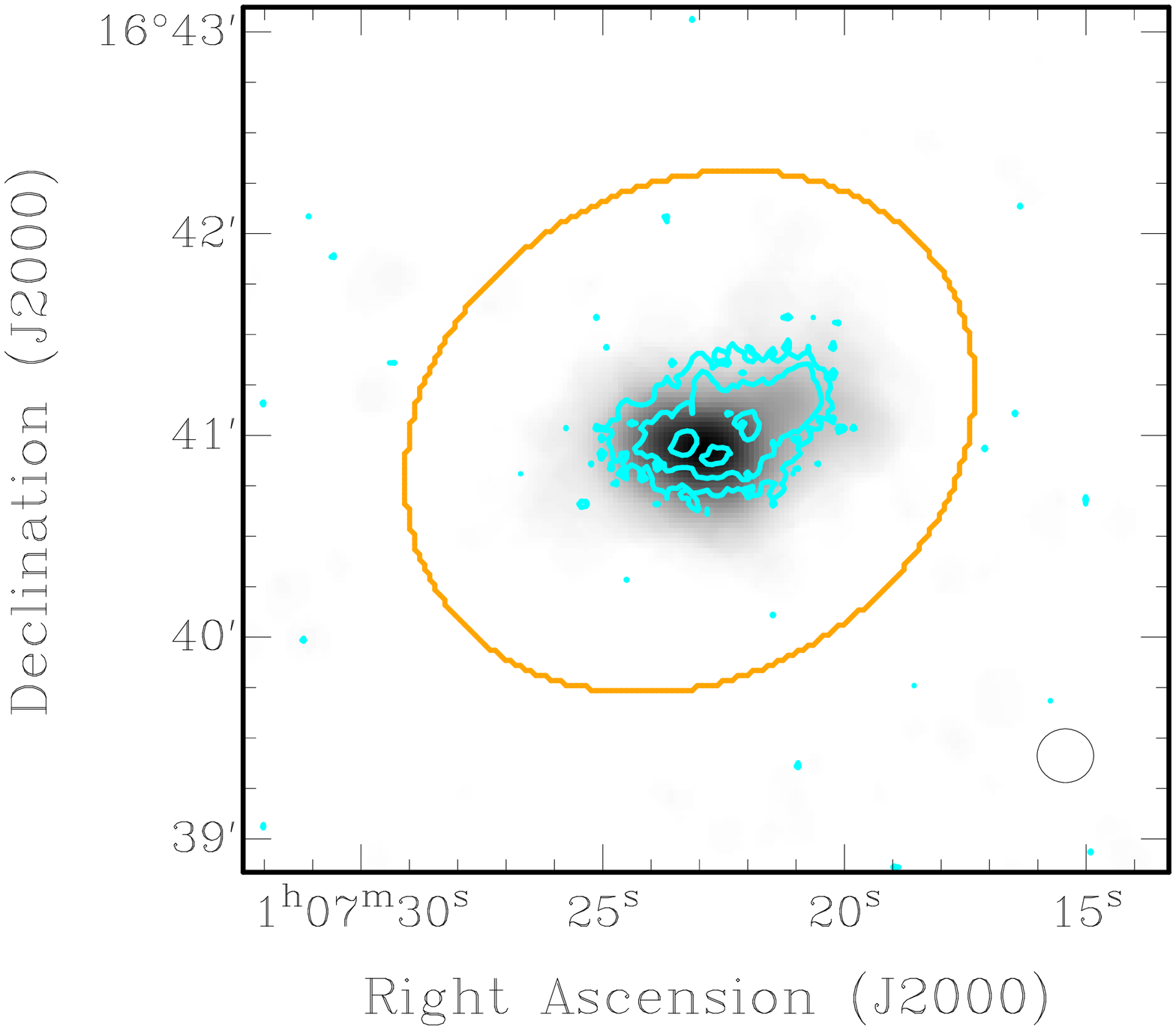}}}&
{\mbox{\includegraphics[height=4cm,angle=270]{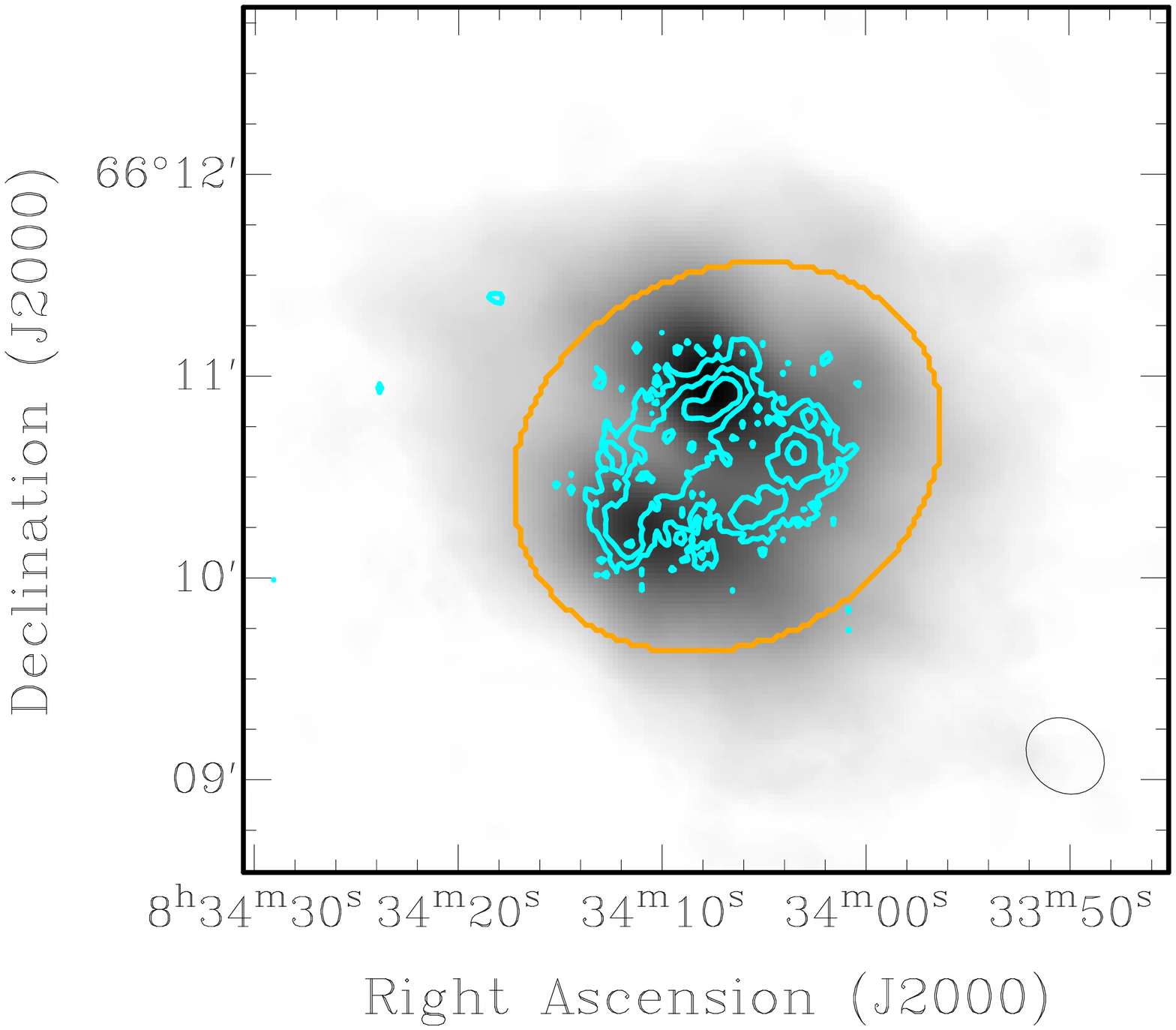}}}&
{\mbox{\includegraphics[height=4cm,angle=270]{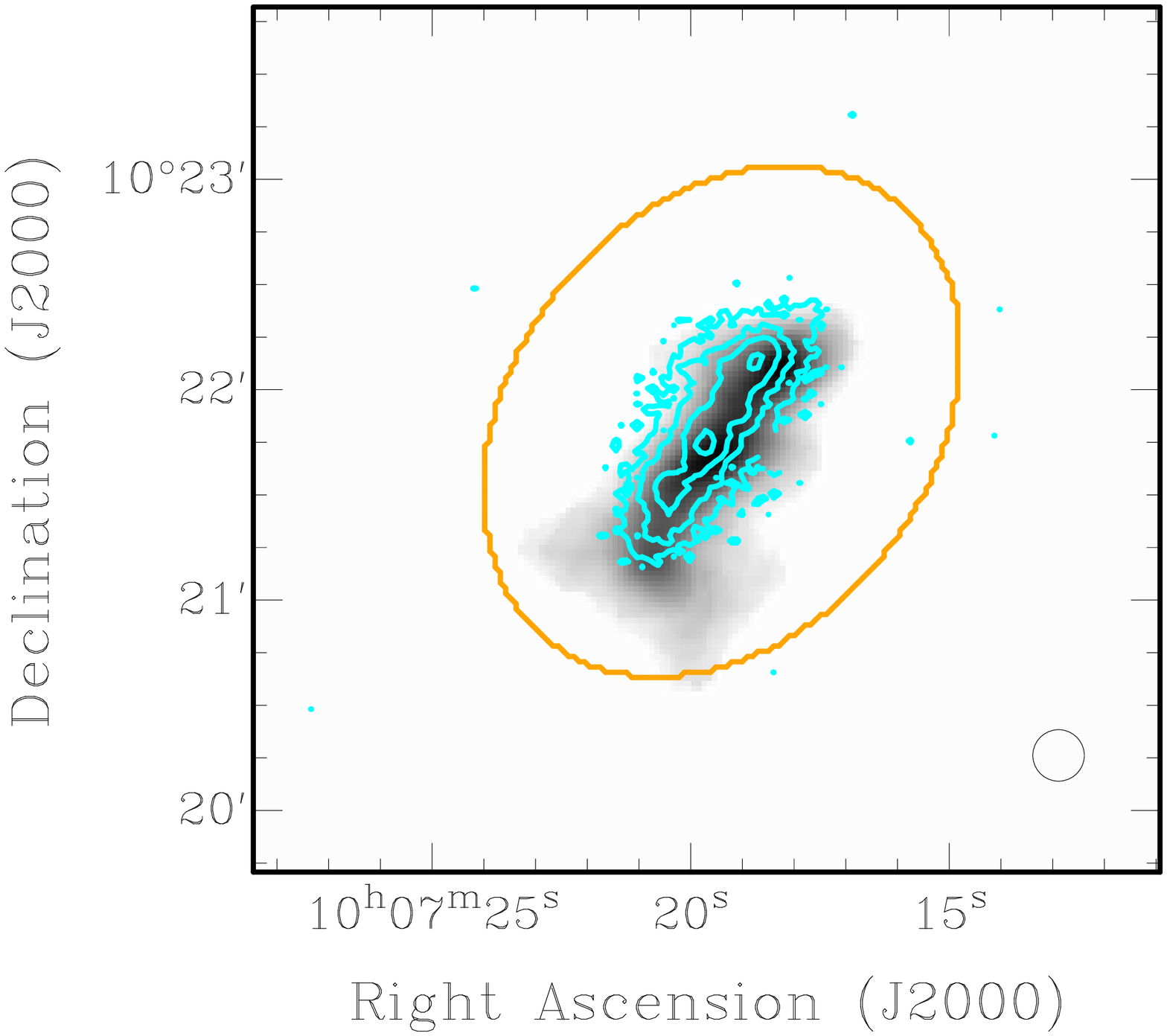}}}\\
\\
NGC 3741&CGCG 269$-$049&DDO 125&UGC 7605\\
{\mbox{\includegraphics[height=4cm,angle=270]{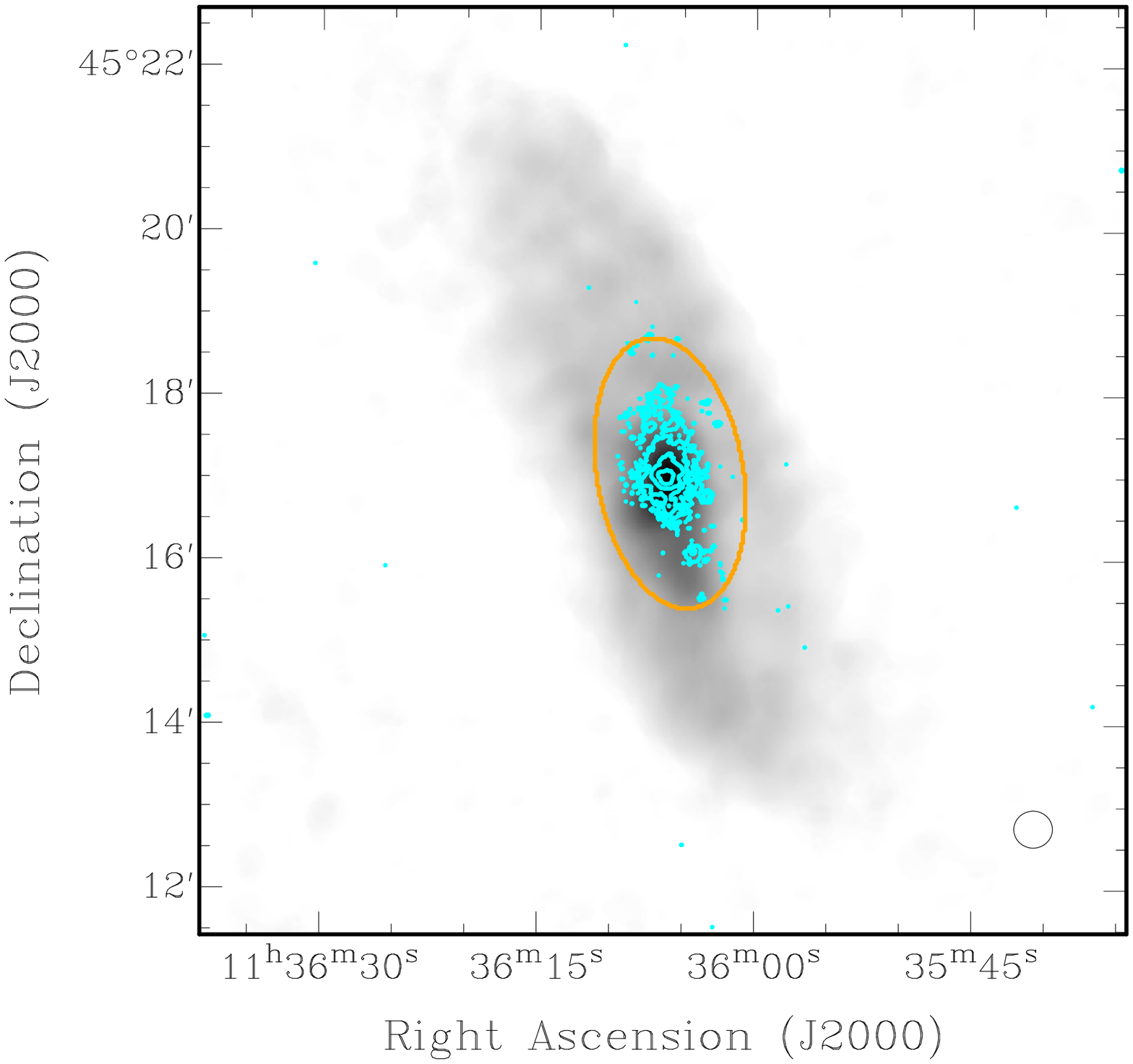}}}&
{\mbox{\includegraphics[height=4cm,angle=270]{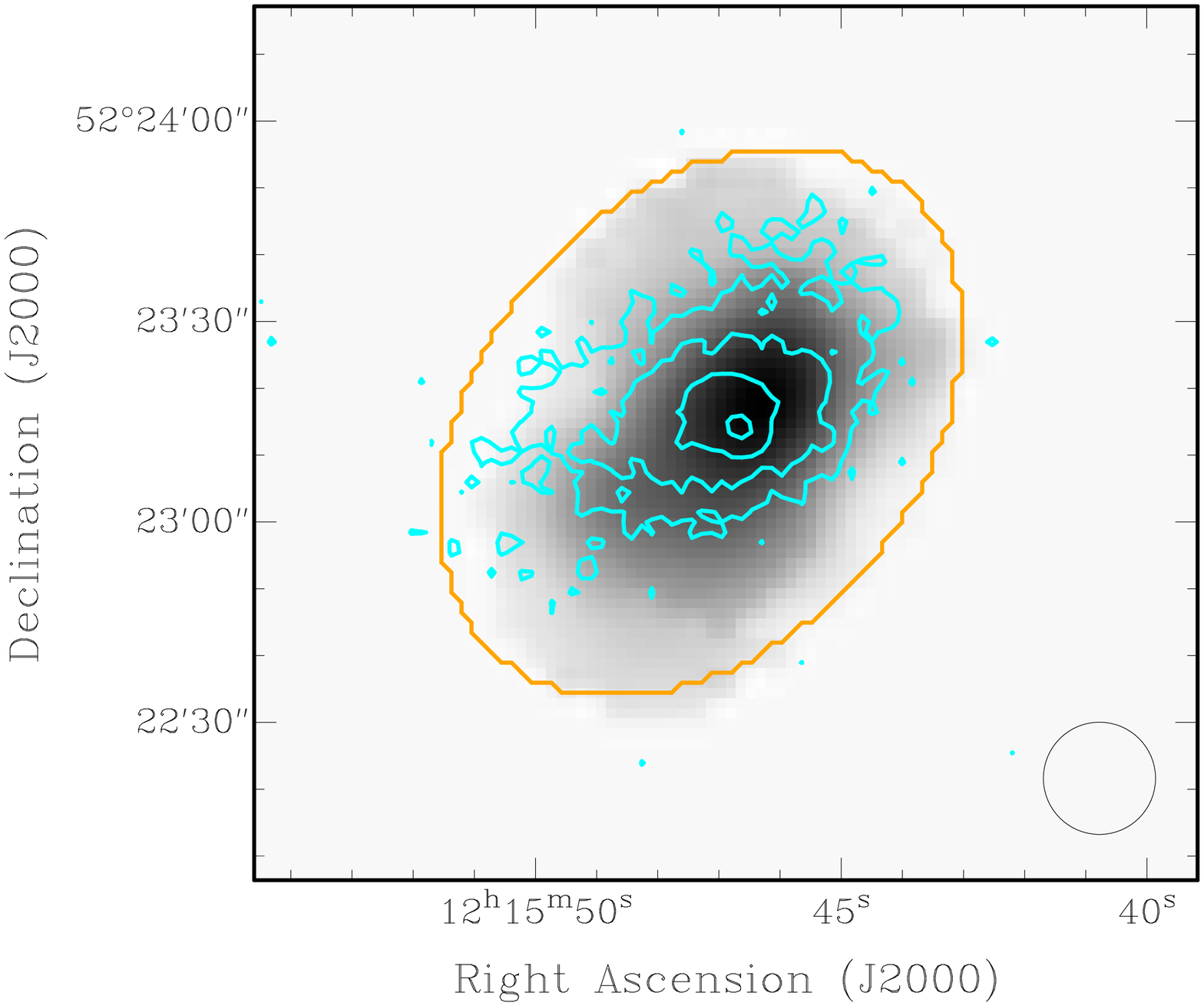}}}&
{\mbox{\includegraphics[height=4cm,angle=270]{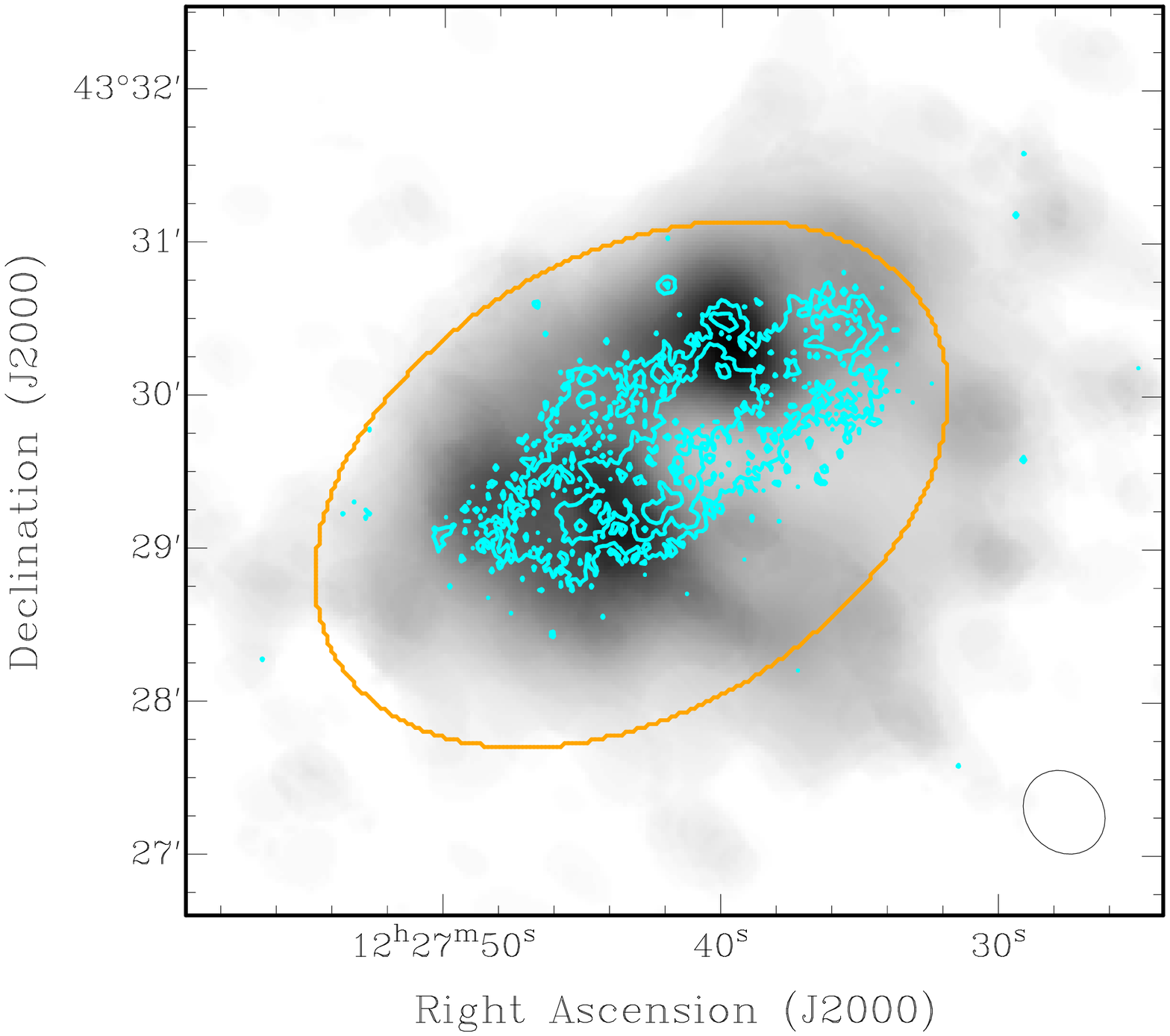}}}&
{\mbox{\includegraphics[height=4cm,angle=270]{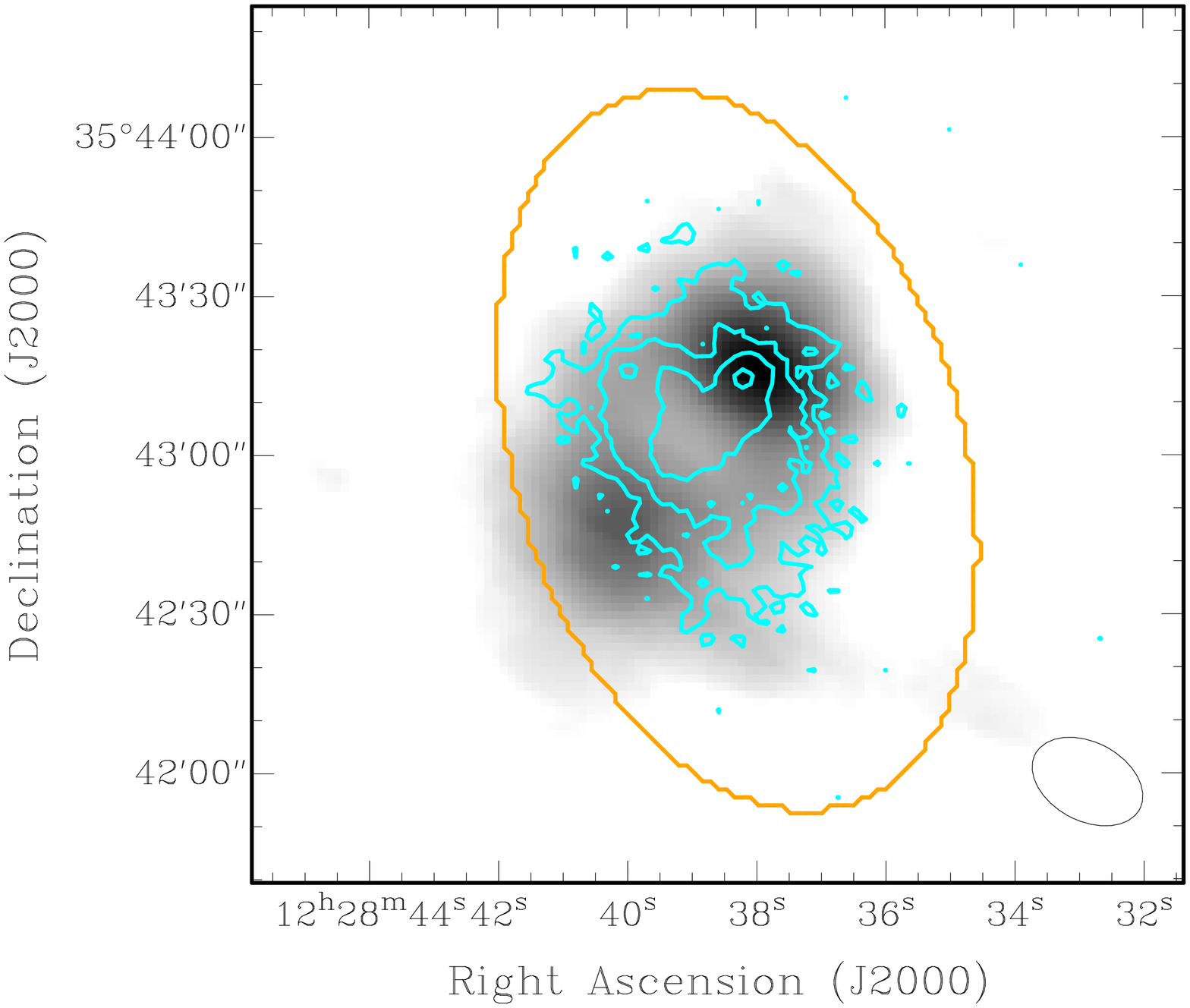}}}\\
\\
GR 8&UGC 8638&DDO 181&DDO 183\\
{\mbox{\includegraphics[height=4cm,angle=270]{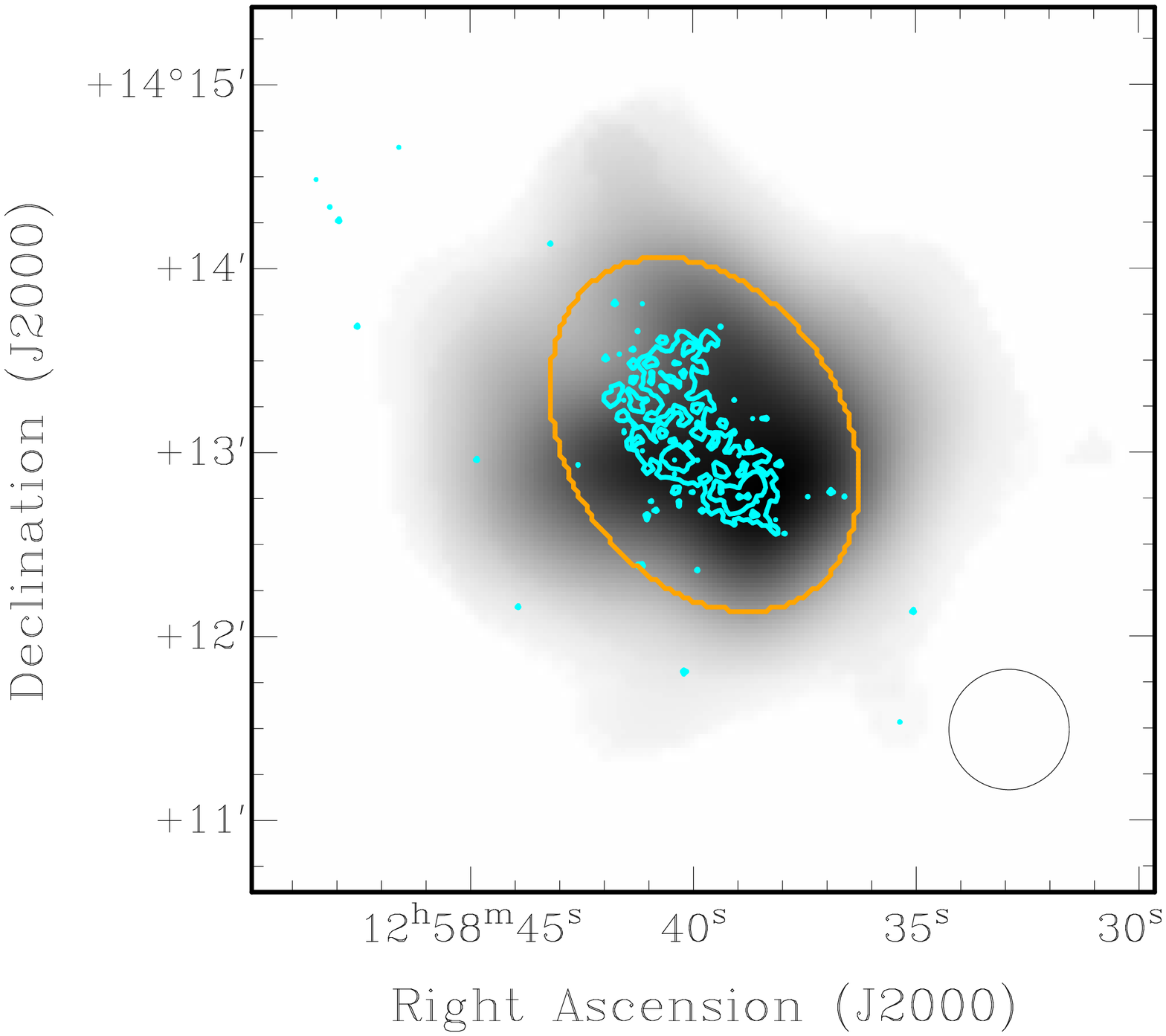}}}&
{\mbox{\includegraphics[height=4cm,angle=270]{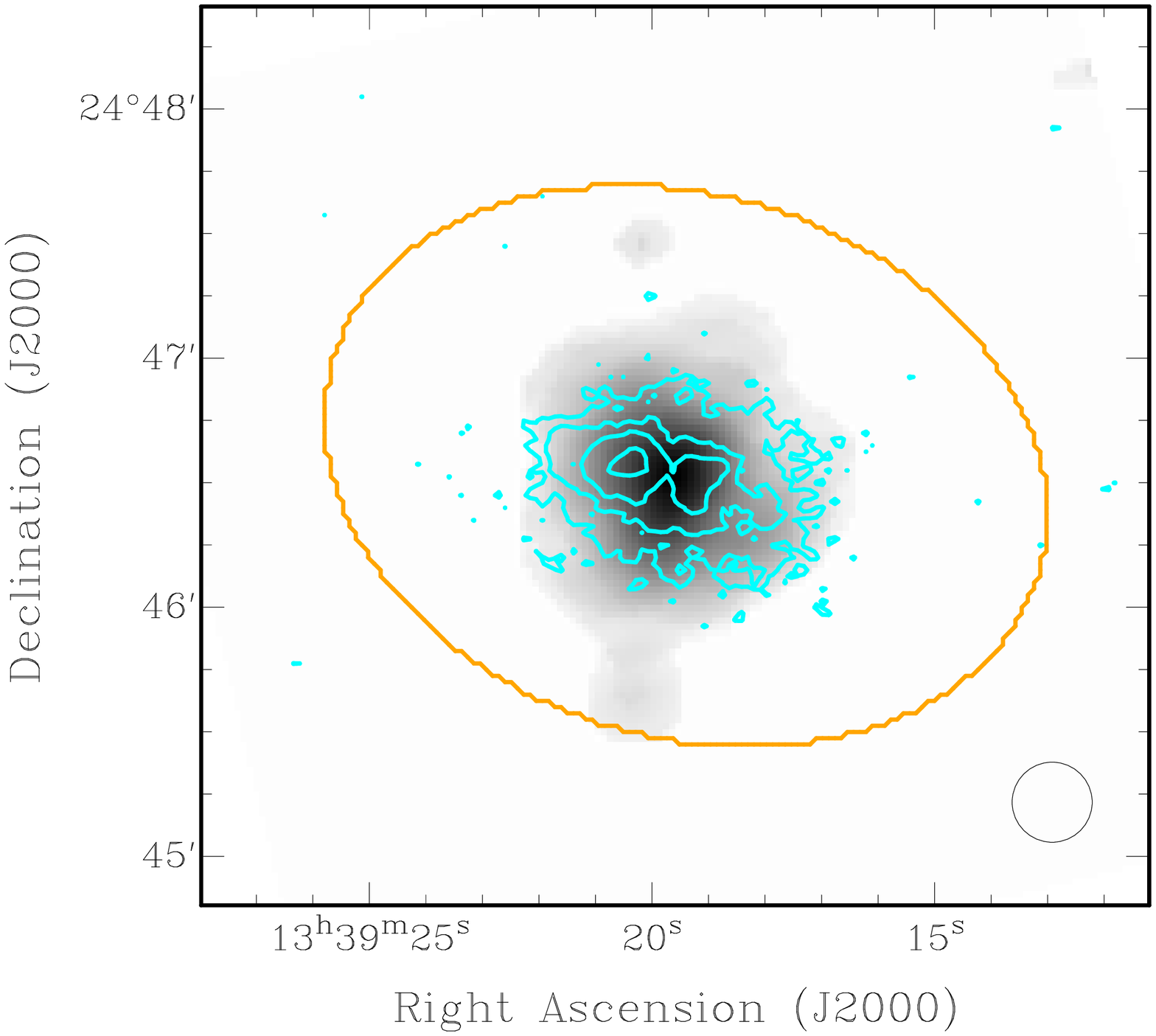}}}&
{\mbox{\includegraphics[height=4cm,angle=270]{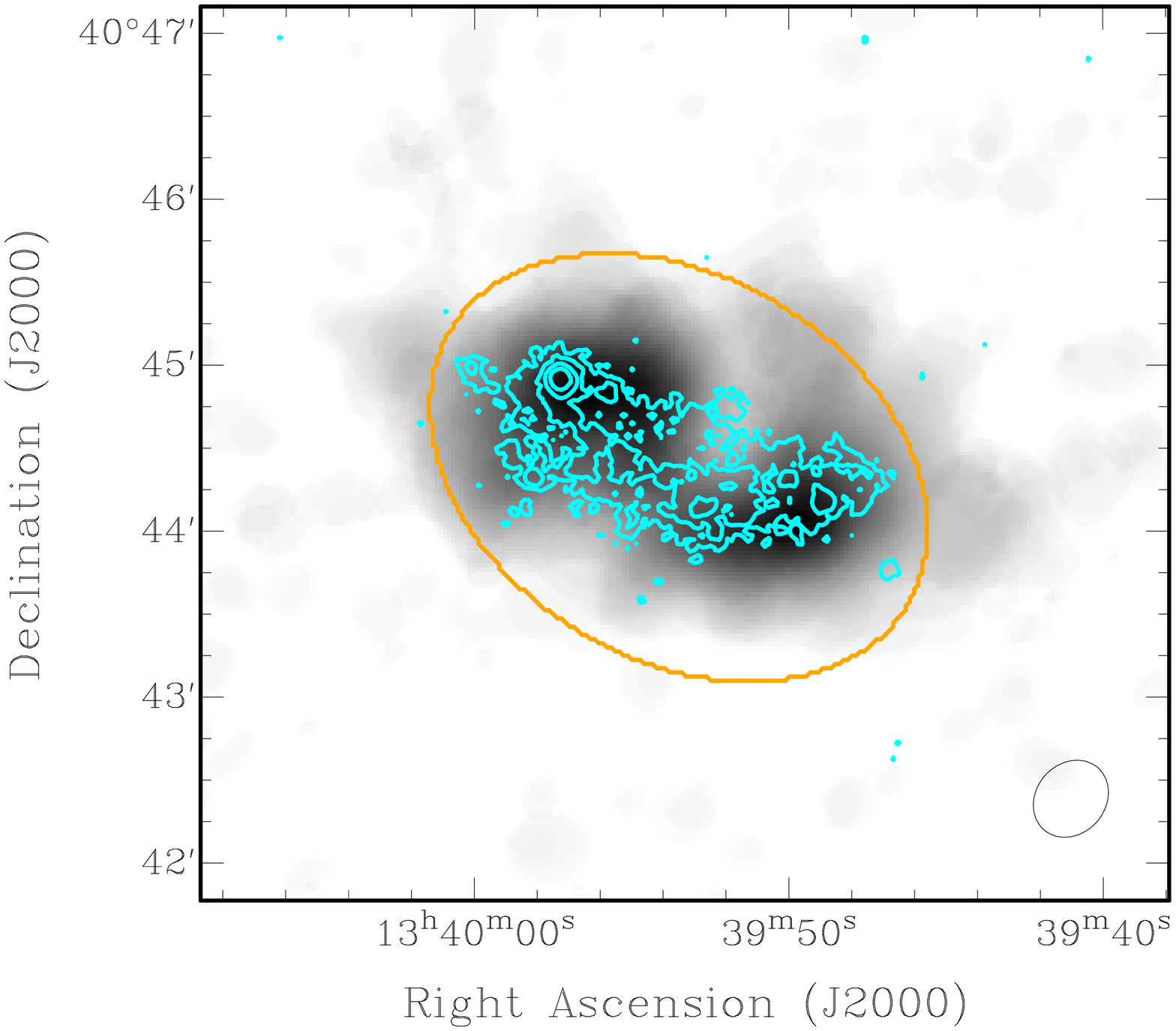}}}&
{\mbox{\includegraphics[height=4cm,angle=270]{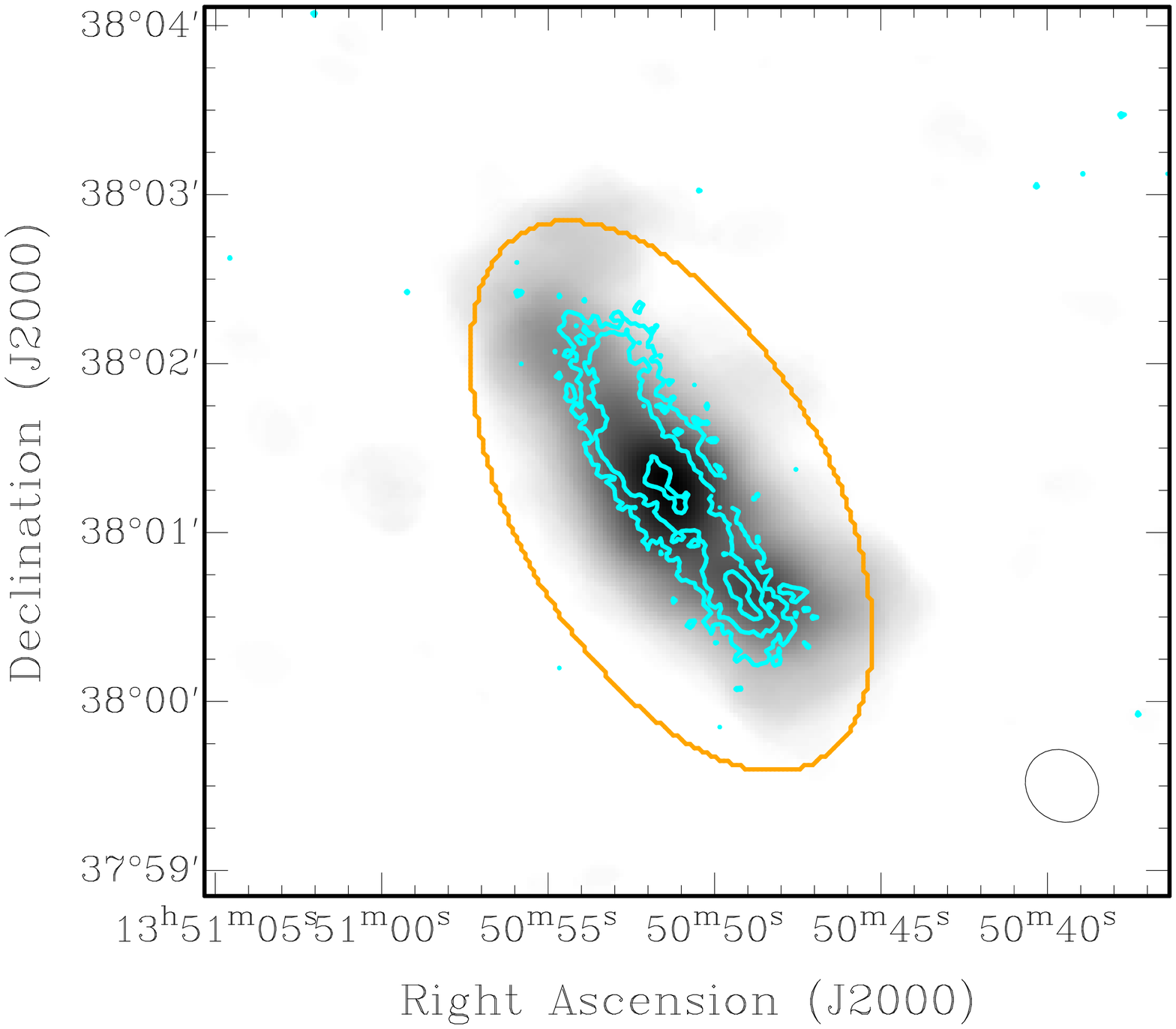}}}\\
\\
UGC 8833&DDO 187&KKH 98&\\
{\mbox{\includegraphics[height=4cm,angle=270]{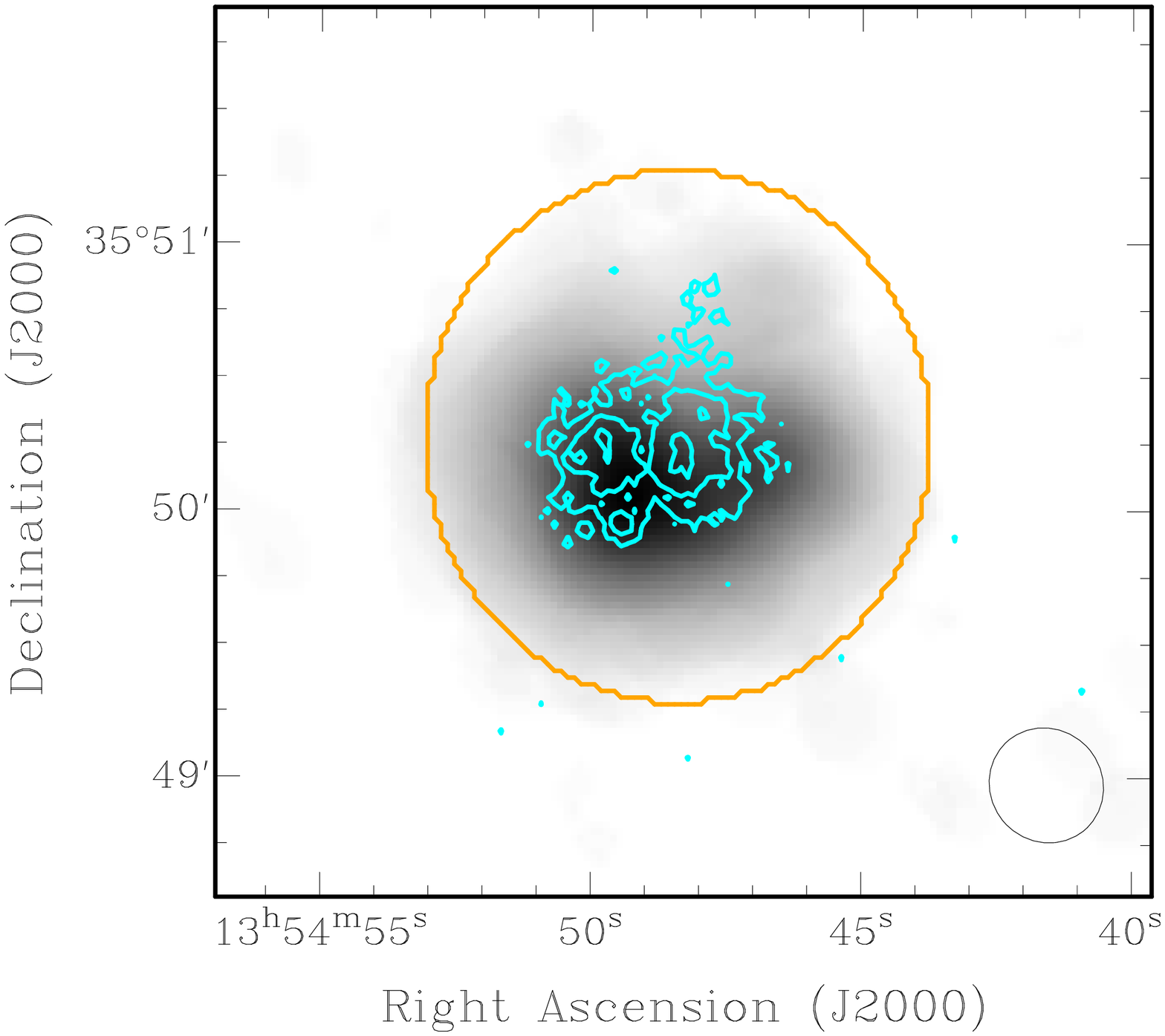}}}&
{\mbox{\includegraphics[height=4cm,angle=270]{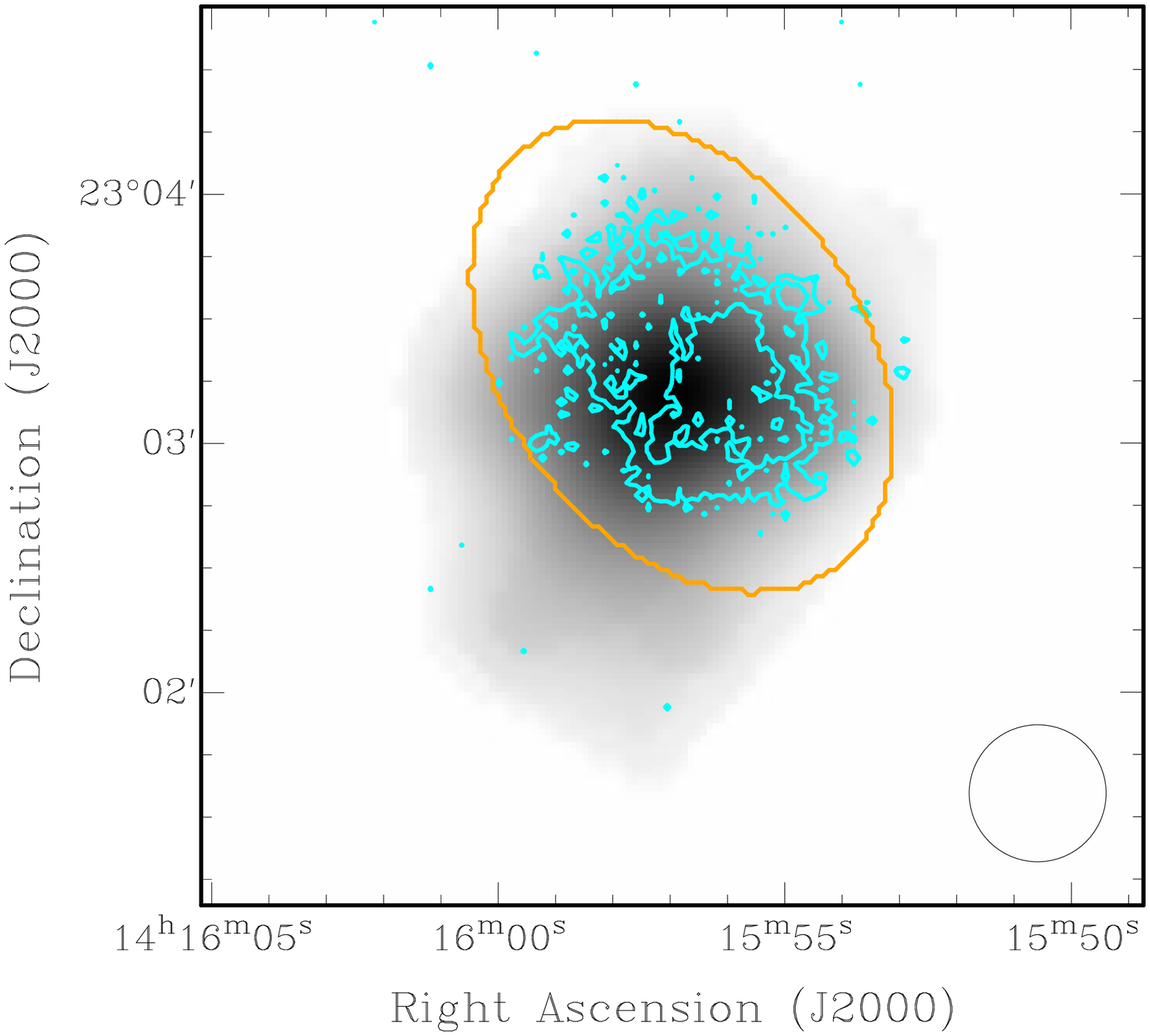}}}&
{\mbox{\includegraphics[height=4cm,angle=270]{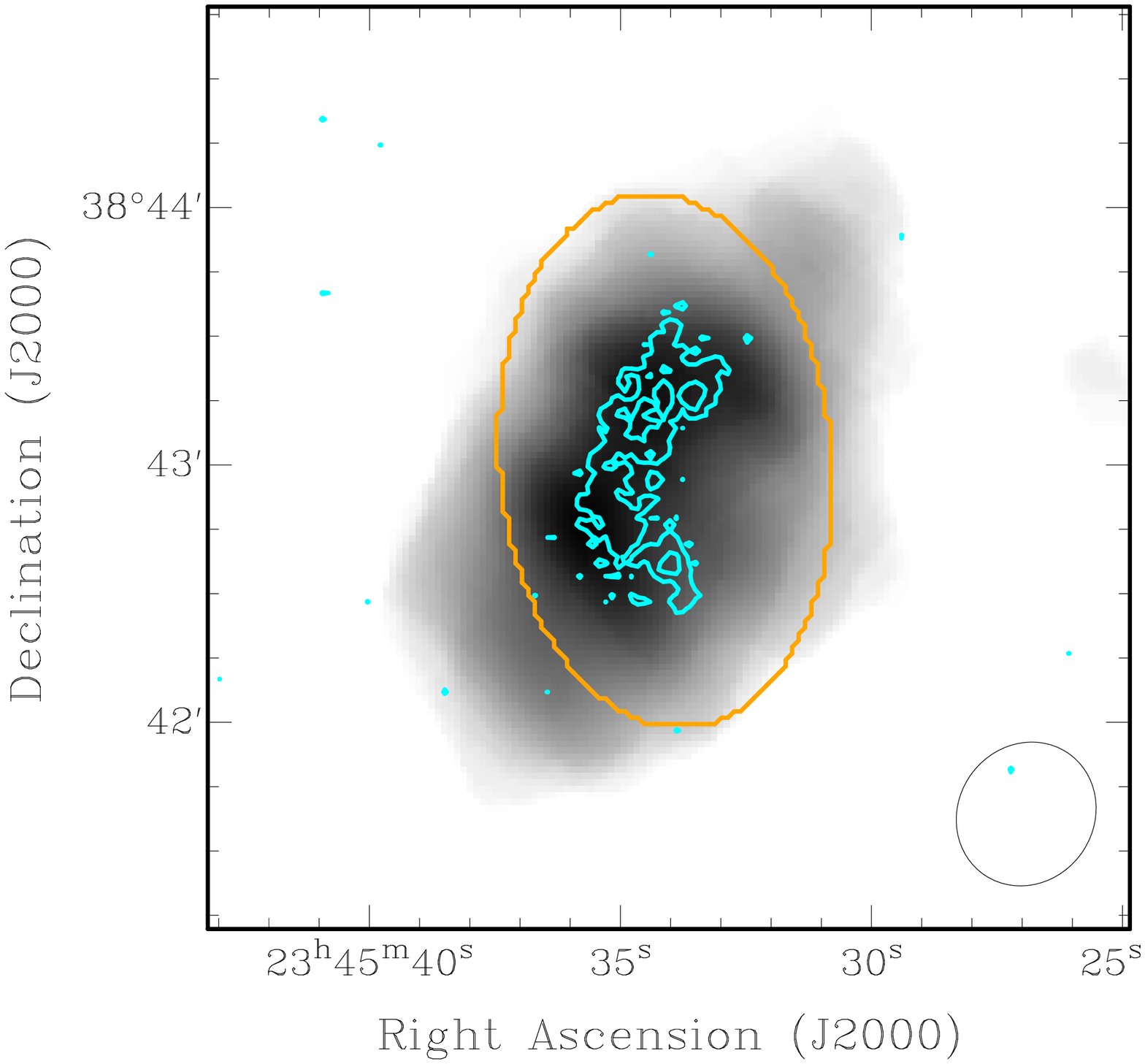}}}&\\
\end{tabular}
\caption{Visual comparison of the extents of HI and star-formation vis-a-vis the stellar disc. FUV flux contours in cyan overlayed on HI column density in greyscale for sample galaxies. The ellipse over which 3.6 $\mu$m flux is extracted in \citet{dal09} is shown in orange. For the FUV emission the level of the first contour is arbitrarily chosen so that traces of background emission are present, and subsequent contours are in multiples of 4. The HI beams are shown at the bottom right corner of each panel.}
\label{fig:ov}
\end{center}
\end{figure*}

Since we aim to compare our results to those found in S11, we measured the surface densities of atomic gas (\shi), SFR and stars within the `stellar disc' - defined to be the isophote within which the 3.6 $\mu$m fluxes were determined, and the same are listed in Table~\ref{tab:samp}.
When determining all the above three quantities, a factor equal to the measured axial ratio of the Holmberg isophote \citep{kar13} is used to correct for the fact that we aim to measure the surface densities perpendicular to the optical disc of the galaxy and that the optical discs of such faint dwarf galaxies are oblate spheroidal in shape \citep{roy13}.
The relative extents of the above mentioned `stellar discs' in relation to the HI and FUV emissions can be seen in Fig.~\ref{fig:ov}.
In all cases the `stellar disc' encompasses the star-forming region as defined by the FUV emission.
But regarding the HI emission, the full range of possible variations is seen viz. the `stellar disc' being less extended, equally extended or more extended than the HI emission.

The HI total intensity maps used to measure the HI column density were constructed to have beam sizes roughly correspond to a linear resolution of 400 pc.
This is a common linear resolution at which all the galaxies from the FIGGS sample can be mapped given their varying distances and the range of baseline lengths available for our GMRT observations.
It is also useful for comparison with earlier studies of the Kennisutt-Schmidt relation which were done using a similar HI resolution \citep[see][for examples]{roy09}.
The HI column density is estimated by summing up the HI flux within the `stellar disc' and using the standard transformation for emission from an optically thin medium.
The value is then multiplied by a factor for 1.36 to account for the presence of helium.
Taking into account the calibration errors we assign a 10\% error on the measured HI column densities.

For calculating the SFR from which \ssfr\ is subsequently calculated, the FUV flux within the `stellar disc' is measured after masking out background and foreground sources.
The SFR is then calculated using the second Equation in Table 3 of \citet{hao11}, which includes a correction for UV flux absorbed by dust and re-radiated at infrared wavelengths.
The correction for dust extinction is done using 24 $\mu$m fluxes from \citet{dal09}, as \citet{dal09} measure these fluxes within the ellipses identical to the `stellar disc' defined above.
24 $\mu$m fluxes are available for all galaxies listed in Table~\ref{tab:samp} except two, viz. UGC 8833 and KKH 98.
For these two galaxies a 24 $\mu$m flux was approximated using eqn. (5) from \citet{roy14}.
The equation gives the best fit relation between the measured 24 $\mu$m flux and the SFRs measured using the FUV flux only, for FIGGS galaxies with both sets of fluxes available.
Finally, a correction factor is applied to account for the low metallicities of the ISM in these galaxies, using values from \citet{rai10} \citep[see][for details]{roy14,roy15}.
We estimate the error on the measured SFRs by adding the following terms in quadrature: measurement error for FUV and 24 \um\ fluxes (when applicable), 10\% flux calibration error for {\it GALEX} FUV data, 5\% flux calibration error for {\it Spitzer} 24 \um\ data, and 50\% error to account for the uncertainty in the SFR calibration caused by variations in the IMF and star formation history \citep{ler12,ler13}.

To calculate $\Sigma_{*}$, we use the 3.6 $\mu$m fluxes in equation C1 from \citet{ler08}. 
They arrive at the mentioned equation by applying an empirical conversion from 3.6 $\mu$m to K-band luminosity combined with a standard K-band mass-to-light ratio.
The major uncertainty comes from the assumed constant mass-to-light ratio which can be as high as 0.2 dex.
Taking the other sources of uncertainty into consideration too, we use a conservative estimate of 0.3 dex error on our measured $\Sigma_{*}$ values.

\section{Results and Discussion}
\label{sec:rnd}

\begin{figure*}
\begin{center}
\begin{tabular}{cc}
{\mbox{\includegraphics[height=8cm]{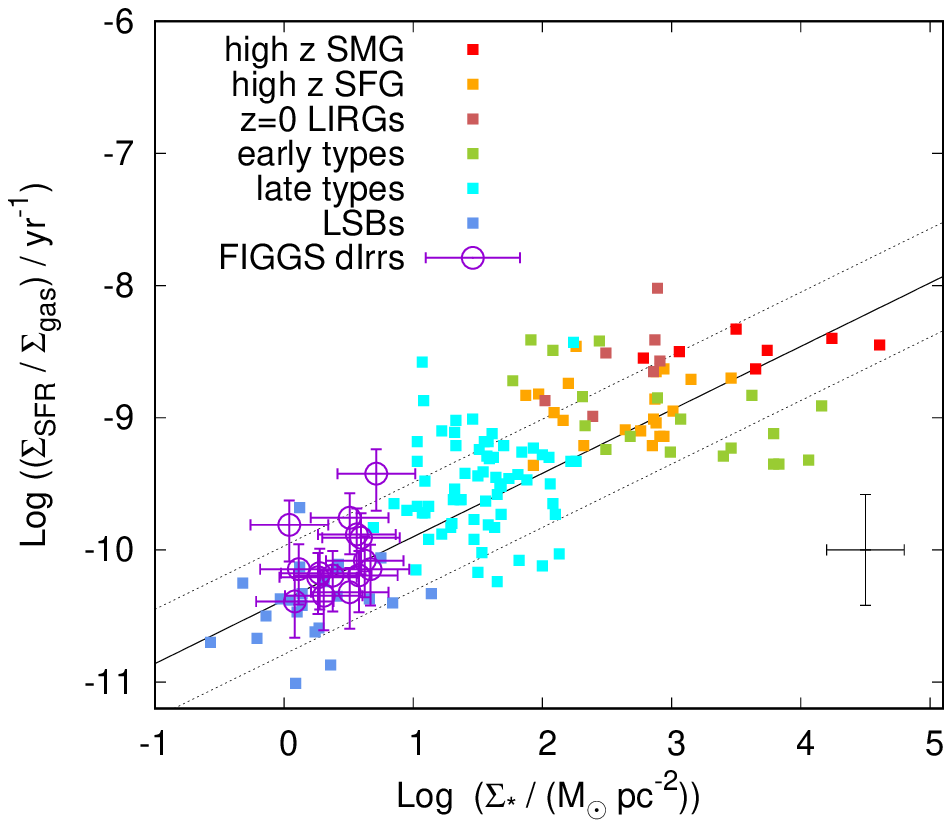}}}&
{\mbox{\includegraphics[height=8cm]{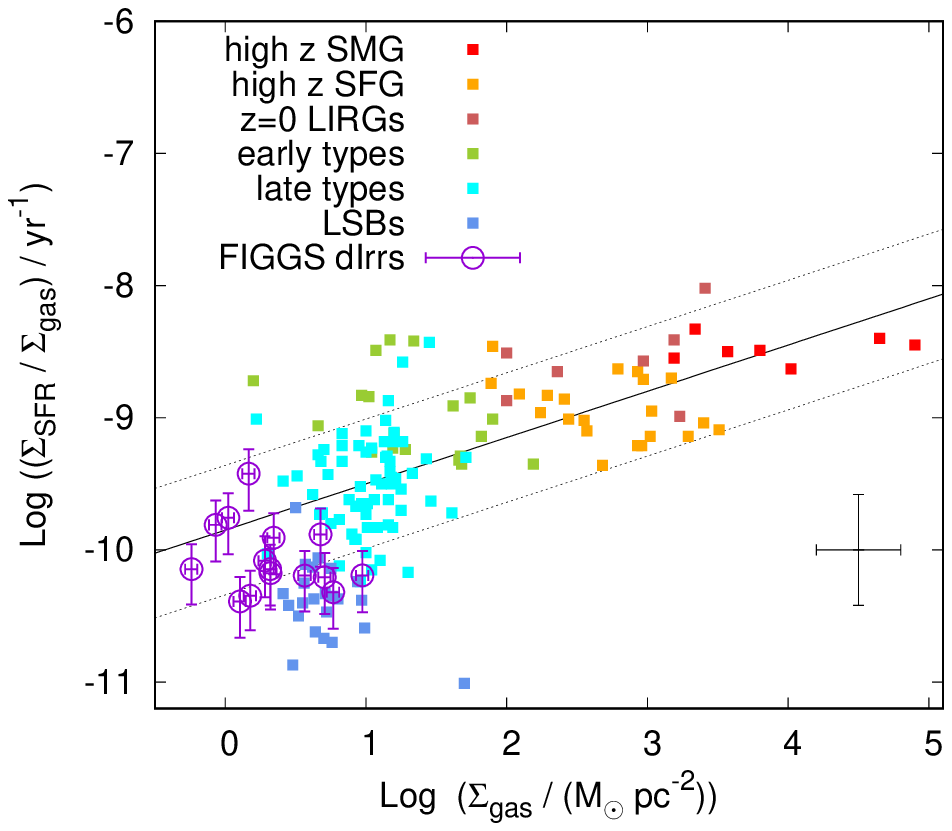}}}\\
\end{tabular}
\end{center}
\caption{{\it Left panel:} the star formation efficiency as a function of stellar mass surface density. The different classes of galaxies from S11 are represented by differently coloured filled squares. The typical errorbars for S11 data points are represented by the black error bars. FIGGS galaxies are represented by violet circles with errorbars. The {\it solid} and {\it dotted} lies are the best fit and 1$\sigma$ scatter around the best fit from S11. {\it Right panel:} the star formation efficiency as a function of gas surface density. All symbols are the same as in left panel. The {\it solid} and {\it dotted} lies are the best fit and 1$\sigma$ scatter around the best fit from S11 excluding early types and LSBs.}
\label{fig:1a}
\end{figure*}

In the left panel of Fig.~\ref{fig:1a} where the star formation efficiency is plotted against $\Sigma_{*}$, our sample galaxies are over-plotted on the ESL relation along with all the galaxies from S11.
Note that the points from S11 have much larger errorbars for \sgas (0.3 dex) as they have to account for the uncertainty in the CO-to-H$_2$ conversion factor.
We clearly see from the plot that dIrrs follow the ESL very well.
Note that as our sample galaxies span a very limited range of $\Sigma_{*}$, we cannot fit a ESL-type relation only to our galaxies and compare such a fit with the ESL from S11.
We do find that our sample galaxies have a mean displacement of only 0.01 dex from the best fit ESL of S11.
Also, our sample galaxies have a rms scatter of 0.21 dex around the best fit relation compared to a rms scatter of 0.41 for the galaxies in S11.
The right panel in Fig.~\ref{fig:1a} where the star formation efficiency is plotted against \sgas, has been included to show that our sample galaxies follow the ESL much more closely than the `canonical' Kennicutt-Schmidt relation.
The right panel shows a variant of the Kennicutt-Schmidt relation with an expected linear slope of 0.4 for the `canonical' relation.
In fact S11 found the slope to be 0.35$\pm$0.04, after excluding the early-type and LSB galaxies which clearly do not follow the same Kennicutt-Schmidt relation as the other galaxies in their sample.
Our sample galaxies have a mean deviation of 0.32 dex from the best fit KS relation by S11 in the right panel.
The galaxies are clearly offset from the best fit relation towards lower star formation efficiencies similar to LSBs.
The lower star formation efficiencies of dIrrs when considering the `canonical' Kennicutt-Schmidt relation has been discussed in our earlier studies \citep{roy09,roy11,roy14}.

In order to explain the physical reason behind the ESL, S11 considered a few possible generic frameworks of star formation.
They found that an ESL-like relationship emerges based on `free-fall in a star dominated potential' for high surface brightness galaxies.
Also for galaxies in which the gravitational effect of the gaseous component can be neglected `pressure-supported star formation' was found to result in power law indices as well as the constants very similar to those in the ESL.
Neither of these models provides an explanation as to why LSB galaxies, which are gas-dominated, should follow the ESL.
It is interesting that the current study indicates that yet another kind of gas dominated galaxy, viz. dIrrs, follow the ESL.

The conclusions in S11 regarding whether pressure-supported star formation {gives an ESL-like relation} is based on an analytical equation of the mid-plane pressure balance from \citet{bli04}, which utilizes that fact that \sst\ traces the mid-plane pressure.
But the said equation from \citet{bli04} is an approximation in that it was based on some necessary but simplistic assumptions.
Therefore we here take a closer look at whether pressure-supported star formation can explain the ESL, given that
there has been a wealth of work since which have built on \citet{bli04}'s ideas of pressure-regulated molecular cloud formation. 
\citet{ost10} proposed a model of star formation based on atomic gas having achieved two-phase thermal and quasi-hydrostatic equilibrium, and later hydrodynamical simulations \citep{kim11,kim13} have shown that these conditions are fulfilled in the atomic gas dominated regions of galaxies when turbulent and thermal pressure are regulated by supernova remnant expansion and stellar FUV heating.
Other analytical laws based on feedback-regulated star formation have been able to explain the variation in KS relation across galaxies too \citep[e.g.][]{dib11}. 
Also recent hydrodynamical n-body simulations \citep{hop14,age15,orr17} have found that feedback on local scales is crucial in setting up the KS law, independent of the recipe used for converting gas to stars, and very mildly dependant on metallicity (more on metallicity below).
All these recent works point to the crucial importance of feedback from earlier generations of stars in setting up the pressure in the interstellar medium and affecting future star formation. And star formation regulated in such a manner can in principle be the reason behind the existence and universality of the ESL, as through the \sst\ term ESL incorporates the effect of previous generations of stars.

For completeness, we next explore the consistency of ESL with an important alternative approach towards relating gas and star formation which uses the fact that metallicity is crucial to setting the atomic-to-molecular transition and thus the SFR \citep{kru09}, and the SFE changes with the metallicity of the system.
The effect is most apparent for low metallicities and at low total gas surface densities \citep[see e.g. Fig. 1 from][]{kru13}, which makes low metallicity galaxies ideal test cases to check the {\it universality} of the model.
In \citet{roy15} we clearly showed that the predictions regarding the KS relation from the \citet{ost10} model better matches the resolved sub-kpc star formation law for FIGGS dIrrs and outer discs of spirals, as compared to predictions from \citet{kru09,kru13}.
Now in this study we have shown that both LSBs and dIrrs follow the ESL, and as discussed above, feedback-regulated, pressure-supported star formation remains the most promising explanation for the ESL.
It would have been ideal to know the exact prediction for the SFE-\sst\ relation based on the feedback/pressure-regulated star formation models and the metallicity-regulated star formation models respectively, and check the predictions against the empirical ESL.
Unfortunately no such predictions exist, and it is beyond the scope of this paper to work them out.
Instead in order to check whether metallicity is a hidden factor behind the ESL, we check whether \sst\ shows any correlation with metallicity.

\begin{table}
\begin{center}
\caption{Average metallicities of galaxies from S11}
\label{tab:met}
\begin{tabular}{|lcl|}
\hline
Galaxy&avg. Z/Z$_{\odot}$&Reference\\
\hline
{\emph{\it late types}}&&\\
NGC0224	&0.69&\citet{pil14}$^1$\\
NGC0598 &0.40&\citet{pil14}\\
NGC0628 &0.62&\citet{pil14}\\
NGC1058 &0.55&\citet{pil14}\\
NGC2336 &0.81&\citet{pil14}\\
NGC2403 &0.33&\citet{pil14}\\
NGC2841 &0.98&\citet{pil14}\\
NGC2976 &0.72&\citet{mou10}$^2$\\
NGC3031 &0.63&\citet{pil14}\\
NGC3077 &0.87&\citet{mar10}$^3$\\
NGC3184 &0.74&\citet{pil14}\\
NGC3198 &0.49&\citet{pil14}\\
NGC3310	&0.42&\citet{pil14}\\
NGC3351 &1.05&\citet{pil14}\\
NGC3368 &2.19&\citet{mar10}\\
NGC3486 &0.49&\citet{pil14}\\
NGC3521 &0.63&\citet{pil14}\\
NGC3627 &0.69&\citet{mou10}\\
NGC3631 &0.63&\citet{pil14}\\
NGC3893 &0.65&\citet{pil14}\\
NGC3938 &0.63&\citet{pil14}\\
NGC4214 &0.35&\citet{mar10}\\
NGC4254 &0.78&\citet{pil14}\\
NGC4258 &0.58&\citet{pil14}\\
NGC4303	&0.62&\citet{pil14}\\
NGC4321 &0.85&\citet{pil14}\\
NGC4449 &0.42&\citet{ber12}$^4$\\
NGC4501 &1.02&\citet{pil14}\\
NGC4535 &0.72&\citet{pil14}\\
NGC4651 &0.56&\citet{pil14}\\
NGC4654 &0.60&\citet{pil14}\\
NGC4713 &0.28&\citet{pil14}\\
NGC4736 &0.69&\citet{pil14}\\
NGC4826 &1.23&\citet{mou10}\\
NGC5033 &0.63&\citet{pil14}\\
NGC5055 &0.85&\citet{pil14}\\
NGC5194	&1.05&\citet{pil14}\\
NGC5236 &0.93&\citet{pil14}\\
NGC5457 &0.39&\citet{pil14}\\
NGC6946 &0.71&\citet{pil14}\\
NGC7331 &0.72&\citet{pil14}\\ 
NGC7793 &0.45&\citet{pil14}\\ 
\hline          
\end{tabular}
\end{center}
\begin{flushleft}
$^1$: 
from metallicity at 0.5~R$_{25}$ using measured O/H abundance gradients, and using 12+log$_{10}$[O/H]$_{\odot}$~=~8.7.\\ 
$^2$: 
from characteristic 12+log$_{10}$[O/H]$_{\rm PT05}$, using 12+log$_{10}$[O/H]$_{\odot}$~=~8.5.\\
$^3$: 
from compiled abundance using 12+log$_{10}$[O/H]$_{\odot}$~=~8.7.\\
$^4$: 
from average abundances using 12+log$_{10}$[O/H]$_{\odot}$~=~8.7.
\end{flushleft}
\end{table}

\addtocounter{table}{-1}
\begin{table}
\begin{center}
\caption{Continued ...}
\label{tab:met}
\begin{tabular}{|lcl|}
\hline
Galaxy&avg. Z/Z$_{\odot}$&Reference\\
\hline
{\emph{\it early types}}&&\\
NGC0524 &1.10&\citet{McD15}$^5$\\
NGC2768	&0.83&\citet{McD15}\\
NGC3032 &0.76&\citet{McD15}\\
NGC3073 &0.18&\citet{McD15}\\
NGC3489 &0.65&\citet{McD15}\\
NGC3773 &0.85&\citet{mou10}\\
NGC4150 &0.52&\citet{McD15}\\
NGC4459 &0.74&\citet{McD15}\\
NGC4477 &0.85&\citet{McD15}\\
NGC4526 &0.89&\citet{McD15}\\
NGC4550 &0.43&\citet{McD15}\\
NGC5173 &0.33&\citet{McD15}\\
UGC9562 &0.26&\citet{pil14}\\
\hline
{\emph{\it LSBs}}&&\\   
DDO154 &0.11&\citet{mou10}\\
HOI    &0.13&\citet{mou10}\\
HOII   &0.17&\citet{mou10}\\
IC2574 &0.22&\citet{mou10}\\
NGC0925&0.52&\citet{mou10}\\
\hline
\end{tabular}
\end{center}
\begin{flushleft}
$^5$: 
from mass-weighted stellar population metallicity measured within R$_e$, using 12+log$_{10}$[O/H]$_{\odot}$~=~8.7.
\end{flushleft}
\end{table}

\begin{figure}
\begin{center}
\includegraphics[height=8cm]{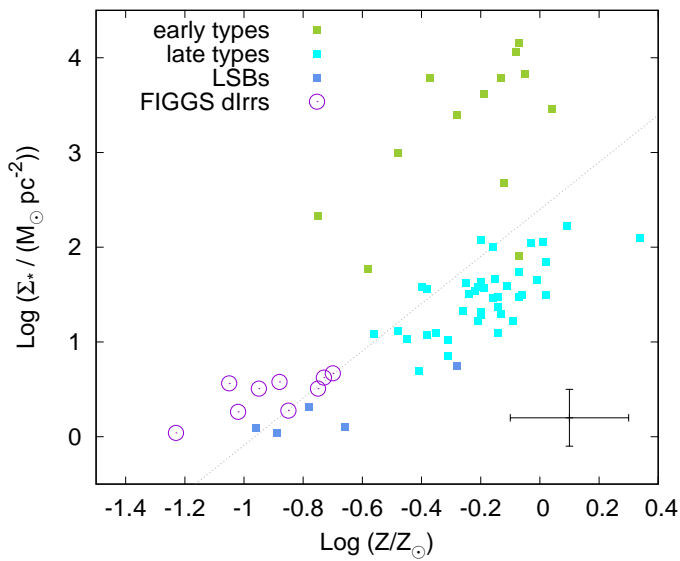}
\end{center}
\caption{The stellar mass surface density ploted against the average metallicity for a sub-sample of S11 and FIGGS galaxies with measured emission metallicities. The different classes of galaxies from S11 are represented by differently coloured filled squares. FIGGS galaxies are represented by violet open circles. The typical errorbars for the data points are represented by the black error bars. The dashed grey line in the background is the best-fit bivariate linear regression line (see text for details).}
\label{fig:Z}
\end{figure}

We use data for galaxies from the S11 sample and from our FIGGS sample with measured emission metallicities, which can provide an indication of the average metallicity over the respective stellar discs. 
We have only considered galaxies from S11 which are classified as `early types', `late types', and `LSBs', as these galaxies along with the FIGGS galaxies cover the low gas surface density end the ESL where the effects of metallicity should be most clearly distinguishable.
For the nine FIGGS galaxies with available emission metallicity measurements, the measurements typically come from the brightest regions of these galaxies.
Therefore these values can be considered as upper limits on the average metallicities over their stellar discs, although it should be noted that dwarf galaxies do not show appreciable metallicity gradients \citep[e.g. see][]{wes13}. 
We also found abundance measurements from the literature for a number of galaxies from the S11 sample which fell into one of the three categories mentioned before.
These are reported in Table~\ref{tab:met} along with the literature reference.
We tried to get an estimate of the average metallicity of the stellar disc of the respective galaxy wherever possible, and also accounted for the fact that the reported metallicities from different sources have different reference solar metallicity levels -- which vary according to the calibration used.  
Wherever needed, we have used the latest measured abundance value for a galaxy.
The relation between \sst\ and the measured metallicities of these galaxies are shown in Fig.~\ref{fig:Z}.
We should note that we are sampling almost the entire range of \sst\ values sampled in the full ESL.
The errors on the measured metallicities mainly come from the solar reference values used.
\citet{kew08} show that the error on the solar reference for any of the standard measures of metallicity based on oxygen emission lines is at most 0.2 dex, a value we conservatively use as the estimate of the error on the measured metallicities.
As we can see from Fig.~\ref{fig:Z}, there is no clear relationship between \sst\ and the average metallicities in the galaxies at the low gas surface density end of the ESL, at least definitely not a linear one.
Each class of galaxies appear to be distinct from each other in the figure, with almost no overlap.
Specifically, dIrrs occupy a distinct region of the \sst--Z space compared to the other three classes of galaxies, especially LSBs which are comparable to them in metallicity.
A bivariate linear regression fit \citep[using the algorithm from][]{kel07}, we get a power law slope of 2.4$\pm$0.2.
More importantly, the algorithm estimates the standard deviation of the intrinsic random scatter of points around the best fit to be 0.73. 
S11 using the same algorithm found the estimated standard deviation of the intrinsic random scatter of points around the best fit ESL relation to be 0.123.

To further confirm that \sst\ and not metallicity is the parameter which drives the star formation law, we follow S11 and try out linear regression fits (IDL regress.pro) in which we treat \ssfr\ as a dependent variable.
The independent variables are (i) \sgas\ and (ii) either of \sst\ or metallicity.
Only using the galaxies with measured metallicities (FIGGS as well as those from the S11 sample), we find the following best-fit power laws:
\begin{equation}
\Sigma_{SFR}~\propto~\Sigma_{gas}^{0.8 \pm 0.1} \Sigma_{*}^{0.33 \pm 0.04},
\end{equation}
\begin{equation}
\Sigma_{SFR}~\propto~\Sigma_{gas}^{1.01 \pm 0.09} (\frac{Z}{Z_{\odot}})^{0.69 \pm 0.12}.
\end{equation}
The difference between the rms scatter of residuals around the respective fits is not statistically significant, but the uncertainty on the exponent when using Z instead of \sst\ is much larger, which reaffirms that \ssfr\ has a much tighter dependence on \sst\  compared to metallicity.
We can therefore conclude that the correlation between SFE and \sst\ (the ESL) is not driven by an underlying correlation between \sst\ and metallicity.

S11 had shown that the \sgas--\sst\ relation for all their sample galaxies is not tighter than the ESL (their Fig. 6), and thus ESL is not a variant of a \ssfr--\sgas\ relation (KS law). 
This study, where low metallicity dIrrs were added to the full sample of galaxies, has helped strengthen the conclusion that ESL is an universally valid fundamental relation defining how gas converts to stars.
Feedback-regulated star formation appears to be a promising candidate to explain the ESL and thus star formation across diverse classes of galaxies.
It is therefore imperative for models incorporating feedback-regulated star formation, or any other model claiming to explain the conversion of gas to stars, to show that they are able to precisely predict the tight correlation that is the extended Schmidt law.

\section*{Acknowledgments} 
We thank Ayesha Begum for providing the visibilities which were used to derive the HI maps for FIGGS galaxies used in this work.
SR thanks Clive Dickinson for his helpful comments.
SR acknowledges support from ERC Starting Grant no. 307209.
YS acknowledges the support from NSFC grant 11373021 and Jiangsu grant BK20150014.


\label{lastpage}

\end{document}